\documentstyle{elsart}
\input epsf
\begin{document}
\begin{frontmatter}
\title{Levels of complexity in financial markets}

\author{Giovanni Bonanno}, \author{Fabrizio Lillo} and \author{Rosario N.
 Mantegna \thanksref{mail1}}

\address{
Istituto Nazionale per la Fisica della Materia, Unit\`a di Palermo\\
and\\
Dipartimento di Fisica e Tecnologie Relative,
Universit\`a di Palermo,\\
Viale delle Scienze, I-90128,
Palermo, Italia}



\thanks[mail1]{corresponding author, e-mail address: mantegna@unipa.it}

\begin{abstract}
We consider different levels of complexity which are observed in 
the empirical investigation of financial time series. We discuss 
recent empirical and theoretical work showing that statistical
properties of financial time series are rather complex under several
ways. Specifically, they are complex with respect to their 
(i) temporal and (ii) ensemble properties. Moreover, the
ensemble return properties show a behavior which is specific to 
the nature of the trading day reflecting if it is a normal or an
extreme trading day.
\end{abstract}


\end{frontmatter}

\section{Introduction}

Financial markets can be regarded as model {\it complex systems}
\cite{Anderson88}. In fact, they are systems composed by many 
agents which are interacting between them in a highly nonlinear 
way. Financial markets are continuously monitored. Data exist 
down to the scale of 
each single communication of bid and ask of a financial asset 
(quotes) and at the level of each transaction (trade). The
availability of this enormous amount of data allows a detailed
statistical description of several aspects of the dynamics of 
asset price in a financial market.  
The results of these studies show the existence of
several levels of complexity in the price dynamics of a financial
asset \cite{MantegnaProc,MS2000,BP2000,Bouchaud2000}. In this 
presentation we will focus on some of them 
that have been investigated by econophysicists and by 
our research group recently.
 
The paper is organized as follows: in the next section we illustrate
the first level of complexity which is observed in the statistical
properties of a single financial time series, Section 3
presents the results obtained by investigating the synchronous
correlations which are observed between all pairs of a selected
set of stocks, Section 4 describes some recent works on the
statistical properties of ensemble return distribution of equities
traded in the New York Stock Exchange and in Section 5 we present 
a brief discussion of our findings.

\section{First level of complexity: time series}

\indent In any financial market---either well established and 
highly active as
the New York Stock Exchange, ``emerging'' as the Budapest stock
exchange, or ``regional'' as the Milan stock exchange---the
autocorrelation function of returns is a monotonic decreasing function
with a very short correlation time.  High frequency data analyses have
shown that correlation times can be as short as a few minutes in highly
traded stocks or indices \cite{Mantegna96,Liu99}. 

This observation is
consistent with the so-called {\it efficient market hypothesis} 
\cite{Samuelson65,Fama}. 
In fact, the short-range memory between returns is directly related to the
necessity of absence of continuous arbitrage opportunities in efficient
financial markets. In other words, the presence of time correlation 
between returns (and then between price changes) would allow devising
trading strategies that would provide a net gain continuously and
without risk. The continuous search for and exploitation of arbitrage
opportunities from traders focused on this kind of activity drastically
diminish the redundancy in the time series of price changes. 

The absence of time correlation between returns does not mean that
returns are identically distributed over time. In fact different authors
have observed that nonlinear functions of return such as the absolute
value or the square are correlated over a time scale much longer than a
trading day.  Moreover the functional form of this correlation seems to
be power-law up to at least 20 trading days approximately
\cite{Liu99,Dacorogna93,Cont97,Cizeau97,Liu97,Pasquini98,Raberto99}.

A final observation concerns the degree of stationary behavior of the 
stock price dynamics. Empirical analysis shows that returns are not 
strictly-sense stationary stochastic processes. Indeed the volatility 
(standard deviation of returns) is itself a stochastic process. Although 
a general proof is still lacking, empirical analyses performed on 
financial data of different financial markets suggest that price
returns and volatility are locally non-stationary but asymptotically 
stationary. By 
asymptotically stationary we mean that the probability density function 
(pdf) of the stochastic variable measured over a wide time interval 
exists and it is 
uniquely defined. A paradigmatic example of simple stochastic processes 
which are locally non-stationary but asymptotically stationary is 
provided by ARCH \cite{Engle82} and GARCH \cite{Bollerslev86} processes.

In summary, the statistical properties of a price time series 
of a financial asset is rather non trivial. The stochastic process
is simultaneously characterized by both short range and long 
range memories and it is stationary only asymptotically. These 
characteristics only would be already enough challenging, 
however, it will not surprise the reader that this is 
only one of several levels of complexity of the price 
dynamics in financial markets.   

\section{Second level of complexity: cross-correlation}

The presence of high degree of cross-correlation between the synchronous time
evolution of a set of equities is a well known empirical fact observed in
financial market \cite{Markowitz59,Elton95,Campbell97}. For a time horizon of
one trading day correlation coefficient as high as 0.7 can be observed for some
pair of equities belonging to the same economic sector.

The study of cross-correlation of a set of economic entities can improve
economic forecasting and  modeling of composed financial entities such as, for
example,  stock portfolios. There are different approaches to address this
problem. The most common one is the principal component analysis of the
correlation matrix of the raw data \cite{Elton71}. This method was 
also used by physicists by using the perspective and theoretical results of the 
random matrix theory \cite{Laloux99,Plerou99}. Another approach
is the correlation based clustering procedure which allow to get 
cluster of stocks homogeneous with respect to the sectors of economic
activities. Different algorithm
exists  to perform cluster analysis in finance \cite{Panton76,Mantegna99,Bernaschi2000}.

Recently \cite{Mantegna99}, it has been proposed to detect economic information
present in a correlation coefficient matrix with a filtering procedure based on
the estimation of the subdominant ultrametric \cite{Rammal86} associated with a
metric distance obtained form the correlation coefficient matrix of set of n
stocks. This method,  already used in other fields, allows to obtain a metric
distance and to extract from it a minimum spanning tree (MST) and 
a hierarchical tree from each
correlation coefficient matrix by means of a well defined algorithm known as
nearest neighbor single linkage clustering \cite{Mardia79}. This allows to
reveal geometrical  (throughout the MST) and taxonomic (throughout the
hierarchical tree) aspects of the correlation present between the stock 
pairs.

In previous work we have shown that this method gives a meaningful taxonomy  for
stock time series \cite{Mantegna99,Bonanno2001} and for market indices of
worldwide stock exchanges \cite{Bonanno2000}. Here we discuss the results
obtained in \cite{Bonanno2001} for stock price time series. The procedure
consists in filtering the relevant information  from the original time series
of returns (i) by  determining the synchronous correlation coefficient of  the
difference of logarithm of stock price computed at  a selected time horizon, 
(ii) by calculating a metric distance  between all the pairs of stocks and (iii)
by selecting the  subdominant ultrametric distance associated to the 
considered metric distance. The subdominant ultrametric is the ultrametric
structure closest to the original metric structure.

The correlation coefficient is defined as 
\begin{equation}
\rho_{ij} (\Delta t) \equiv \frac{<Y_i Y_j>-<Y_i><Y_j>}
{\sqrt{(<Y_i^2>-<Y_i>^2)(<Y_j^2>-<Y_j>^2)}}
\end{equation}
where $i$ and $j$ are  numerical labels of the stocks, 
$Y_i=\ln P_i(t)-\ln P_i(t-\Delta t)$, $P_i(t)$ is the value of 
the stock price $i$ at the trading time $t$ and $\Delta t$ is the
time horizon which is, in the present discussion, one trading day. 
The correlation coefficient 
for logarithm price differences (which almost coincides with 
stock returns) is 
computed between all the possible pairs of stocks present in 
the considered portfolio. 
The empirical statistical average, indicated in this paper 
with the symbol $<.>$,
is here a temporal average always
performed over the investigated time period. 

By definition, $\rho_{ij} (\Delta t)$ can vary from -1 (completely 
anti-correlated pair of stocks) to 1 (completely correlated 
pair of stocks).
When $\rho_{ij} (\Delta t)=0$ the two stocks are uncorrelated. 
The matrix of correlation coefficient is a symmetric matrix with 
$\rho_{ii}(\Delta t)=1$ in the main diagonal. 
Hence for each value of $\Delta t$, $n~(n-1)/2$ correlation 
coefficients characterize 
each correlation coefficient matrix completely.

A metric distance between pair of stocks can be rigorously 
determined \cite{Gower66} by defining
\begin{equation}
d_{i,j} (\Delta t)=\sqrt{2(1-\rho_{ij}(\Delta t))} .
\end{equation}
With this choice $d_{i,j}(\Delta t)$ fulfills the three axioms of a metric --
(i) $d_{i,j}(\Delta t)=0$ if and only if $i=j$; 
(ii) $d_{i,j}(\Delta t)=d_{j,i} (\Delta t)$ and (iii)
$d_{i,j} (\Delta t) \le d_{i,k} (\Delta t) +d_{k,j} (\Delta t)$. 
The distance matrix ${\bf{D}} (\Delta t)$ is 
then used to determine the MST connecting the $n$ stocks. 

The MST, a theoretical concept of graph theory 
\cite{West96}, is a graph with $n-1$ links which selects 
the most relevant connections of 
each element of the set. The MST allows to obtain, 
in a direct and essentially unique way, the subdominant 
ultrametric distance matrix ${\bf{D}}^<(\Delta t)$ 
and the hierarchical organization of the elements 
(stocks in our case) of the investigated data set.

The subdominant ultrametric distance between $i$ and $j$
objects, i.e. the element $d^<_{i,j}$ of the ${\bf{D}}^<(\Delta t)$  
matrix, is the maximum value 
of the metric distance $d_{k,l}$ detected by moving 
in single steps from $i$ to $j$ through the path connecting 
$i$ and $j$ in the MST. 
The method of constructing a MST linking a set of $n$ objects 
is direct and it is known in multivariate analysis as the
nearest neighbor single linkage cluster analysis \cite{Mardia79}. 
A pedagogical exposition of the determination of the
MST in the contest of financial time series is provided in ref. 
\cite{MS2000}.

Subdominant ultrametric space \cite{Rammal86} has been 
fruitfully used in the description of frustrated complex systems. 
The archetype of this kind of systems is a spin glass \cite{Mezard87}.

In ref. \cite{Bonanno2001}, we investigate a set of 100 highly 
capitalized stocks traded in 
the major US equity markets during the period January 1995 - 
December 1998. At that time, most of them were  used to compute the
Standard and Poor's 100 index. The prices are transaction prices 
stored in the {\it Trade and Quote} database of the 
New York Stock Exchange. 

The time horizons investigated in the cited study 
varies from $\Delta t=d=6$ h and $30$ min (a trading day time interval), 
to $\Delta t=d/20=19$ min and $30$ sec.
Here, we only discuss the case of the one day time horizon to
present the simplest aspect of this kind of complexity detected in
the synchronous dynamics of price returns.

\begin{figure}[t]
\epsfxsize=4.0in
\centerline{\epsfbox{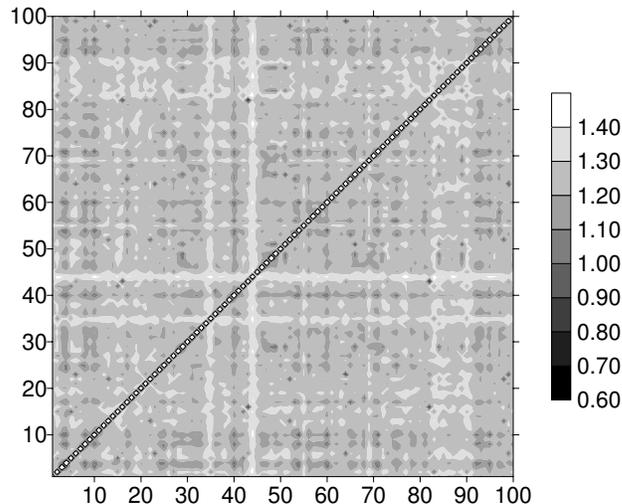}}
\caption{Gray scale table of the distance matrix of our portfolio.
Equities are ordered in alphabetical order from left to right and
from bottom to top. Each gray spot indicates the distance 
$d_{i,j}$ between stock $i$ and stock $j$. The gray scale used
is shown at the right side of the figure. No simple pattern is
detected in the distance matrix.}
\label{fig1}
\end{figure}

In Fig. 1 we show the distance matrix obtained from the
correlation matrix for the $\Delta t=6$ h and $30$ min 
(one trading day) time horizon. 
The figure uses a gray scale to indicate the distance between 
each pair of equities . The order of the equities in the 
rows and columns of the distance matrix is the alphabetical
order of their tick symbols. Of course this order has no 
economic meaning associated with it. By using this ordering
procedure, the distance matrix does
not show a simple recognizable pattern so that one cannot conclude 
about the presence or absence of relevant information in it. 

\begin{figure}[t]
\epsfxsize=4.0in
\centerline{\epsfbox{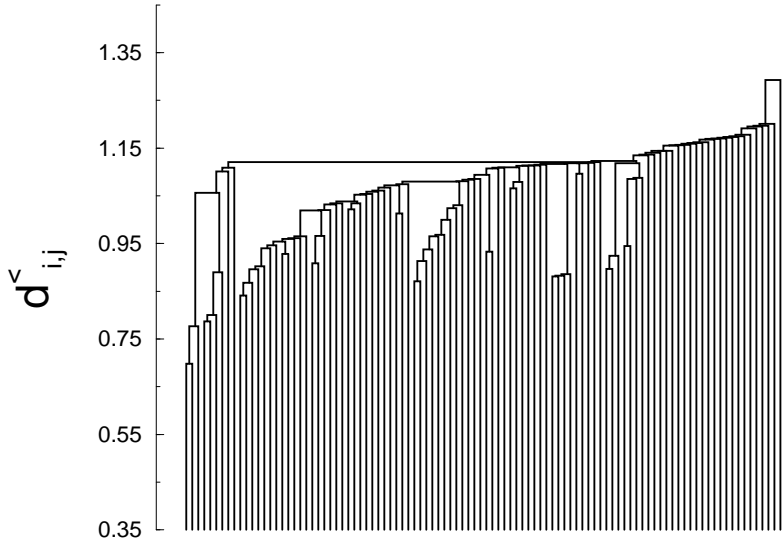}}
\caption{Hierarchical tree of the set of 100 stocks traded in the
US equities markets obtained starting from the return time series 
computed with a
$\Delta t=6$ h and 30 min time horizon (1 trading day) during the time period
Jan 1995-Dec 1998. Each stock is indicated by a vertical line.
The hierarchical tree is highly structured.
Two stocks (lines) links when a horizontal line is drawn between
two vertical lines. The height of the horizontal line indicates the 
ultrametric distance at which the two stocks are connected.
The tick simbols of the investigated stocks are from left to right:
SLB,                                               
HAL,                                               
BHI,                                               
MOB,                                               
CHV,                                               
XON,                                               
ARC,                                               
OXY,                                               
CGP,                                               
JPM,                                               
BAC,                                               
MER,                                               
ONE,                                               
WFC,                                               
AXP,                                               
AIG,                                               
KO,                                               
GE,                                                
PG,                                                
CL,                                                
USB,                                               
MRK,                                               
BMY,                                               
JNJ,                                               
AGC,                                               
CPB,                                               
PEP,                                              
WMT,                                               
MAY,                                               
S,                                                 
DIS,                                               
CI,                                                
UTX,                                               
MCD,                                               
NT,                                                
NSC,                                               
BNI,                                               
RAL,                                               
MSFT,                                              
INTC,                                              
TXN,                                               
CSCO,                                              
SUNW,                                              
NSM,                                              
IBM,                                               
HWP,                                               
ORCL,                                              
AVP,                                               
HON,                                               
BAX,                                               
GM,                                                
F,                                                 
BA,                                                
HRS,                                               
DOW,                                               
DD,                                                
MMM,                                               
IFF,                                               
HNZ,                                               
VO,                                                
XRX,                                               
WY,                                               
CHA,                                               
IP,                                                
BCC,                                               
FDX,                                               
DAL,                                               
LTD,                                               
AA,                                                
CSC,                                               
BEL,                                               
AIT,                                               
GTE,                                               
SO,                                                
AEP,                                               
UCM,                                               
ETR,                                               
GD,                                                
MTC,                                               
MO,                                                
ROK,                                               
TAN,                                               
PNU,                                               
WMB,                                               
BDK,                                               
TOY,                                               
MKG,                                               
RTNB,                                              
CEN,                                               
EK,                                                
PRD,                                               
UIS,                                               
TEK,                                               
BS,                                                
T,                                                 
COL,                                               
FLR,                                               
BC,                                                
KM,                                                
HM. For a description of the investigated stocks see a financial
web site as, for example, www.quicken.com.}
\label{fig2}
\end{figure}

The correlation coefficient matrix and the directly related 
distance matrix are not of straightforward interpretation if 
no processing of the information contained in them has been 
performed.

We process this information by extracting from the distance matrix the
ultrametric distance matrix associated with it.  The knowledge of the
ultrametric distance matrix allows to obtain a hierarchical tree (a taxonomy)
without the  use of any external threshold or clustering parameter. The
hierarchical tree obtained starting from the distance matrix  described in Fig.
1 is shown in Fig. 2. In the figure,  each vertical line indicates  an equity.
Each of the investigated stocks is indicated with its tick symbol in the figure
caption. Several clusters are  clearly identified. From left to right, the most
prominent  are (i) the cluster  of energy stocks (SLB, HAL, BHI, MOB, CHV, XON,
ARC, OXY and CGP), (ii) the cluster of financial stocks (JPM, BAC, MER, ONE,
WFC, AXP and AIG),  (iii)  technology cluster (MSFT, INTC, TXN, CSCO, SUNW,
NSM, IBM, HWP and ORCL),  (iv) basic materials cluster (WY, CHA, IP and BCC)
and  (v) utility cluster (BEL, AIT, GTE, SO, AEP, UCM and ETR). 

The direct interpretation of the economic clusters of Fig. 2 
shows that a set of one-day time horizon 
time series of returns carries
information about the economic sector of the stocks considered.

\begin{figure}[t]
\epsfxsize=4.0in
\centerline{\epsfbox{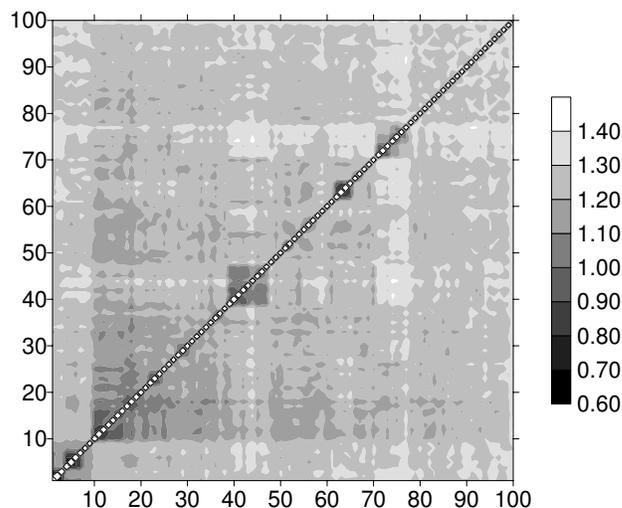}}
\caption{Gray scale table of the distance matrix of our portfolio.
Equities are ordered with the order obtained from the 
hierarchical tree of Fig. 2 (from left to right and
from bottom to top). Each gray spot indicates the distance 
$d_{i,j}$ between stock $i$ and stock $j$. The gray scale used
is the same as in Fig. 1. Clusters are clearly observable with 
the present ordering. From the left to the right prominent 
examples of clusters are the ones of oil (from 1 to 9) , 
financial, conglomerates and consumer/non-cyclical 
(from 10 to 21), technology (from 39 to 47), 
basic materials (from 62 to 65) and
utility (from 71 to 76).}
\label{fig3}
\end{figure}

The hierarchical tree provides an order of portfolio equities
that can be used to rearrange the distance matrix. By using this
order obtained by investigating the subdominant ultrametric 
of the distance matrix we plot the distance matrix in a form which 
is much more readable than in the case shown in Fig. 1. In fact
in Fig. 3 we observe the presence of groups of stocks which form
clusters (to black areas in the matrix) and we also observe
regions of longer distance (to white regions in the matrix). 

Several clusters are directly observable in the distance matrix 
builded up with the present ordering. From left to right prominent 
examples of clusters are the ones of (i) oil
which is, to be precise, a cluster composed by two separated 
sub-clusters to which belongs the companies SLB, HAL, BHI 
(companies which are providing financial services to 
the oil industry) and MOB, CHV, XON, ARC, 
OXY, CGP (company of the oil and gas industry); (ii) 
financial (JPM, BAC, MER, BAC, ONE, WFC, APX, etc)
and consumer/non-cyclical companies (KO, GE, PG, CL, JNJ, etc);
(iii) technology companies (MSFT, INTC, TXN, CSCO, NSM, IBM, HWP,
ORCL); (d) basic materials (paper industry WY, CHA, IP, BCC)
and (iv) utility companies (BEL, AIT, GTE, SO, AEP, UCM, ETR). It 
may be worth noting that financial and consumers/non-cyclical
companies are characterized by short distances also with some companies 
which are located outside 
their cluster whereas the distances of oil, technology and paper
companies with companies outside their cluster are usually 
homogeneously much longer than distances inside their 
clusters. This effect is most evident for the utility cluster.

Equity time series are then carrying economic information which 
can be detected by using specialized filtering procedures as the one
we discuss in the present paragraph. In summary, price time series 
in a financial market are not only complex in their 
time statistical properties but they are also rather complex 
with respect to the intricate synchronous interaction of each 
time series with all the others.

\section{Third level of complexity: Collective behavior during 
extreme market events}

In the present paragraph we discuss a third level of complexity.
Specifically, we discuss the different statistical behavior 
observed in a set of equities simultaneously traded in
a financial market during typical and extreme market days.  

The investigation of the return distribution of an ensemble 
of stocks simultaneously traded in a financial market 
was introduced in \cite{Lillo99}. 
The statistical properties of price return distribution of an 
ensemble of stocks are discussed in \cite{Lillo2000} for the 
typical trading days and in \cite{Lillo2000b} for the extreme
market days.

In both studies, the investigated market is the New York Stock 
Exchange (NYSE) during the 12-year period from January 1987 
to December 1998 which corresponds to 3032
trading days. The total number of assets $n$ traded in NYSE is 
rapidly increasing and it ranges from $1128$ in 1987 to 
$2788$ in 1998. The total number of data records exceeds $6$ million. 

The variable investigated in our analysis is the daily price return, 
which is defined as $R_i(t)\equiv (P_i(t+1)-P_i(t))/P_i(t)$,
where $P_i(t)$ is the closure price of $i-$th asset at day $t$ ($t=1,2,..$). 
In our studies, we 
consider only the trading days and we remove the weekends and the holidays 
from the data set. Moreover we do not consider price returns which are 
in absolute values greater than $50\%$ because some of these returns might 
be attributed to errors in the database and may affect in a considerable 
way the statistical analyses. We extract the $n$ returns of the 
$n$ stocks for each trading day and we consider the normalized pdf 
of price returns. 
The distribution of these returns gives an idea about the 
general activity of the market at the selected 
trading day. In the absence of extreme events, the central part of 
the distribution is conserved for long time periods.
In these periods the shape of the distribution is systematically 
non-Gaussian and approximately symmetrical \cite{Lillo2000}. During 
extreme trading days the pdf changes abruptly its shape either 
in the presence of positive or negative mean return. 

\begin{figure}[t]
\epsfxsize=4.0in
\centerline{\epsfbox{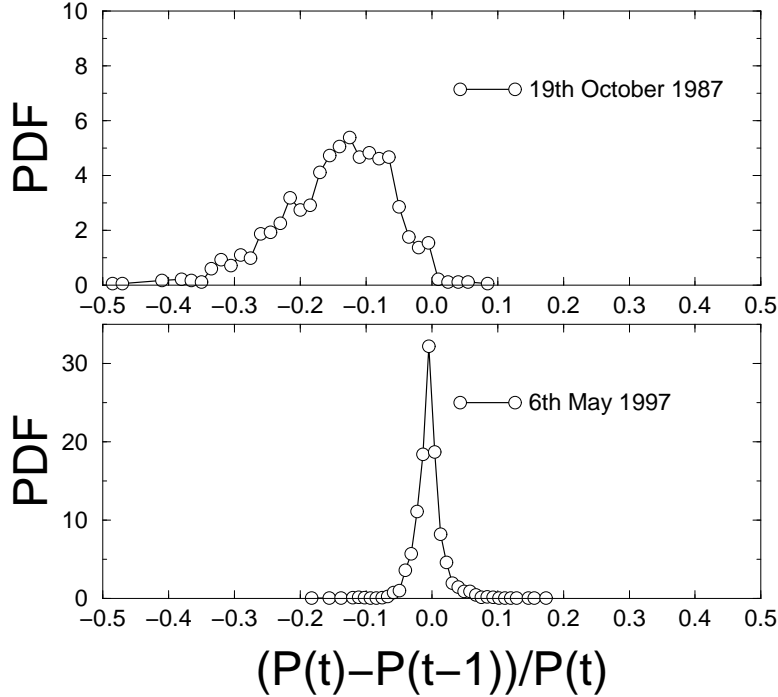}}
\caption{Daily ensemble return distribution of all the equities 
traded in the New York Stock Exchange for the trading days: 19th October
1987 (top) and 6th May 1997 (bottom). The 19th October 1987 is 
the Black Monday, the worst trading day in financial markets since 
last 50 years whereas the 6th May 1997 is a typical trading day. 
The shape of the ensemble distribution dramatically changes from 
typical (bottom) to extreme (top) market days.}
\label{fig4}
\end{figure}

In Fig. 4 we show the daily ensemble return distribution of all the equities 
traded in the New York Stock Exchange for two representative trading days,
the famous 19th October
1987 (top) and the anonymous 6th May 1997. 
The 19th October 1987 is the Black Monday,
the worst trading day in financial markets since last 50 years whereas
the 6th May 1997 is a typical trading day. The shape of the ensemble
distribution dramatically changes from typical (bottom) to extreme
(top) market days. The most surprising change in the shape of 
the ensemble return distribution concerns its symmetry property.
In ref. \cite{Lillo2000b}, one shows that the shape of the ensemble return
distribution is symmetrical in the typical trading day whereas 
in extreme days (crashes or rallies) the distribution
is skewed (negatively or positively respectively). 
 A quantitative estimate 
of the asymmetry of the pdf is difficult in finite statistical sets 
because the skewness parameter depends on the third moment of 
the distribution. Moments higher than the second are essentially 
affected by rare events rather than by the central part of the  
distribution. Due to the finite number of stocks in our statistical 
ensemble, a measure of the asymmetry of the distribution based 
on its skewness is not statistically robust. 

We overcome this problem in ref. \cite{Lillo2000b} 
by considering an alternative measure of 
the asymmetry of the distribution. Specifically, we extract the 
median and the mean of the distribution for all 
trading days. When a probability distribution 
is symmetric the median coincides with the mean. Therefore the
difference between the mean and the median is a measure
of the degree of asymmetry of the distribution. For 
positively (negatively) skewed distribution the median is smaller 
(greater) than the mean. The median depends weakly on the 
rare events of the random variable and therefore is much less 
affected than the skewness by the finiteness of the number 
of records of the ensemble.

\begin{figure}[t]
\epsfxsize=4.0in
\centerline{\epsfbox{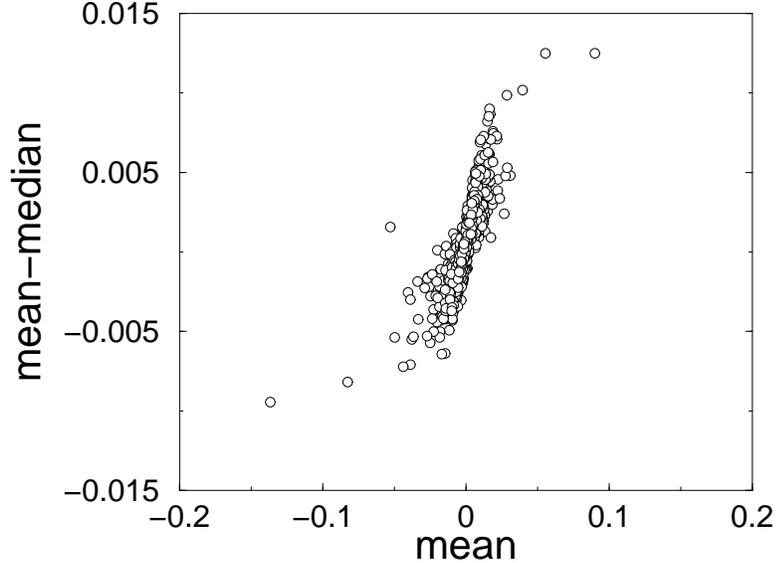}}
\caption{Degree of symmetry of the ensemble 
return distribution quantified by the difference 
between the mean and the median 
of the ensemble return distribution as a function of the mean 
for each trading day. Each circle refers to a single trading day
of the investigated time period. The change in asymmetry is detected 
both for crashes (left region of the figure) and for rallies
(right region of the figure). Typical and extreme market events are 
collapsing over a sigmoid function.}
\label{fig5}
\end{figure}
      
Figure 5 shows the difference between the mean and the median as 
a function of the mean for each trading day of the investigated period.
In the Figure each circle refers to a different trading day.
The circles cluster in a pattern which has a
sigmoid shape. In days in which the mean is positive (negative) 
the difference between mean and median is positive (negative). 
In extreme days, the 
corresponding circles are characterized by a great absolute value of 
the mean and a great absolute value of the difference between mean 
and median. 
Another result summarized in 
Fig. 5 is that symmetry alteration is not exclusive of the days of extreme 
crash or rally but it is also evident for trading days 
of intermediate absolute mean return.
The change of the shape and of the symmetry properties during 
the days of large absolute returns suggests that in extreme days 
the behavior of the market cannot be statistically described 
in the same way of the 'normal' periods. Moreover Figure 5 indicates
that the difference from normal to extreme behavior 
increases gradually with the absolute value of the average 
return.

In ref. \cite{Lillo2000,Lillo2001} we compare the results of
empirical analysis discussed above 
with the ones predicted by a simple model: the 
single-index model. The single-index 
model \cite{Elton95,Campbell97} assumes that the returns of all 
assets are controlled by one factor, usually called the market. 
For any asset $i$, we have
\begin{equation}
R_i(t)=\alpha_i+\beta_i R_M(t)+\epsilon_i(t),
\end{equation}
where $R_i(t)$ and $R_M(t)$ are the return of the asset $i$ and of the 
market at day $t$, respectively, $\alpha_i$ and $\beta_i$ are two 
real parameters and $\epsilon_i(t)$ is a zero mean noise term 
characterized by a variance equal to $\sigma^2_{\epsilon_i}$.  
The noise terms of different assets are uncorrelated, 
$<\epsilon_i(t) \epsilon_j(t)>=0$ for $i\neq j$. Moreover 
the covariance between $R_M(t)$ and $\epsilon_i(t)$ is set to
zero for any $i$.

Each asset is correlated with the market and the presence of such a 
correlation induces a correlation between any pair of assets. 
It is customary to adopt a broad-based stock index for the market $R_M(t)$. 
Our choice for the market is the Standard and Poor's 500 index.
The best estimation of the model parameters $\alpha_i$, $\beta_i$
and $\sigma^2_{\epsilon_i}$ is usually done with the ordinary least 
squares method \cite{Campbell97}. In our comparison 
\cite{Lillo2000,Lillo2001}, we infer that  
the correlation which are detected among the stocks can be described 
by the single-index model only as a first approximation. The 
degree of approximation of the single-index model progressively
becomes worst for market days of increasing absolute average return
and fails in properly describing the market 
behavior of extreme days. Discrepancies between the theoretical 
predictions of the single-index model and empirical data have 
also been documented in ref. \cite{Cizeau2001}.

In summary, a third level of complexity is also present in financial
markets. Typical and extreme days are different with respect to the 
statistical properties of the ensemble return distribution. Specifically,
in addition to the first moment (mean return) also higher moments 
governing the shape of the ensemble return distribution change during
extreme (crashes or rallies) market events.
However, the change of shape of the ensemble return distribution 
is not arbitrary and statistical regularities can be detected for 
extreme events occurring after a time interval as long as ten years.

\section{Discussion}

A complete modeling of the dynamics of financial markets 
turns out to be extremely challenging. Several levels of
complexity arise from the investigation of statistical
properties of a set of price of financial assets simultaneously
traded in a market. Each single time series has statistical
properties which are rather difficult to model (in fact a 
definitive model is still waiting in spite of all the
efforts devoted to this problem since the seminal paper of
Bachelier). In addition to this
level of complexity, each set of financial assets  
has associated a specific overall dynamics which connotes economic
information driving the global system. The nature and dynamical 
properties of the correlations between stocks are a key
aspect of the complexity of a financial market. The degree
of this complexity is enhanced by the observation that
the market behave in a different way in typical and
in extreme days.

Pointing out all these levels of complexity may sound 
frustrating for researchers interested in modeling 
such a system. However, we wish to point out that there
may be another interpretation of these results: 
the results obtained in making formal the 
existence of these levels of complexity suggest that 
the system eventually obeys some deep rules that control
the statistical properties of the global system both
in typical and in extreme days. We believe this is indeed 
the case that makes so challenging and interesting the study of
such {\it complex systems}.

\begin{ack}

The authors thank INFM and MURST for financial support. This work 
is part of the FRA-INFM project 'Volatility in financial markets'. 
G. Bonanno and F. Lillo acknowledge FSE-INFM for their fellowships.

\end{ack}

\end{document}